# Reply to "Comment on 'Quantum Phase Shift Caused by Spatial Confinement'" by Murray Peshkin

B.E. Allman, A. Cimmino and A.G. Klein
School of Physics, University of Melbourne, Victoria 3010, Australia

We describe a force-free phase shift due only to temporal geometric boundary conditions placed on a neutron deBroglie wave packet.

1.  INTRODUCTION

Our paper [1] describes the phase shift due only to geometric boundary conditions placed on a deBroglie wave constricted in directions transverse to its motion. The paper begins with a quantum mechanical derivation of the phase shift and then presents the results of two experimentally analogous phase shifts achieved using optical photons. The phase shift experienced by the photons passing through such a constriction is the familiar change in phase velocity of an electromagnetic wave inside a waveguide. The paper also proposes a neutron interferometry experiment to observe this phase shift.

2.  STATIC NEUTRON EXPERIMENT

In the neutron interferometric experiment described in the paper the neutron in one arm of the interferometer enters a one-dimensional channel of wall separation, a, and length, l. It was shown that the longitudinal momentum of the neutron would change when passing through the constriction. This change in longitudinal momentum is given by

$$\Delta p \approx \frac{\pi^2 \hbar^2}{2a^2 p} = \frac{h \lambda}{8a^2}, \qquad (1)$$

and leads to a phase shift given by

$$\Delta \phi \approx \frac{\pi}{4} \frac{\lambda l}{a^2} = \frac{\pi^2 \hbar l}{2a^2 \sqrt{2mE}}. \qquad (2)$$

Peshkin [2] has correctly pointed out that upon entering such a static constriction the neutron would experience a force in the direction of motion due to the spatial gradient of the constriction (nuclear) potential. The force is responsible for the change in longitudinal momentum and results in a dispersive phase shift (eqn.(2)). We point out that such a change in longitudinal momentum could be measured interferometrically but also in a time-of-flight experiment.

3. TEMPORAL NEUTRON EXPERIMENT

A different situation occurs for a temporally modulated constriction rather than a static one. Experimentally, this requires the neutron to be inside the ends of the channel walls when the constriction is formed and then the walls again separate before the neutron reaches their other ends. Such a temporal experiment was originally described by Greenberger [3]. If the wall separation changes from $\infty \to a$, the energy of the particle is changed by

$$\Delta E = \left(\frac{\pi\hbar}{a}\right)^2 \frac{1}{2m}. \qquad (3)$$

Note that the longitudinal momentum (kinetic energy) remains constant, indicative of a force free interaction. Now the phase shift is given by

$$\Delta\phi = \int \frac{\Delta E}{\hbar} dt = \frac{\pi^2 \hbar T}{2a^2 m}, \qquad (4)$$

where T is the effective duration of the constriction. This phase shift is independent of the neutron energy. Such an experiment is a purely geometric analogue of the force-free, non-dispersive Aharonov-Bohm phase shifts [4]. The neutron passing through the temporal channel in a barrier is the analogous consideration to the scalar Aharonov-Bohm experiment [5] wherein the neutron passes through a temporal magnetic field to achieve a non-dispersive phase shift [6]. The above equation may be written as

$$\Delta\phi = \frac{\pi^2 \hbar}{2a^2} \frac{T}{m} = \frac{\pi^2 \hbar}{2a^2} \frac{1}{p} = \frac{\pi\lambda l}{4a^2}, \qquad (5)$$

identical to eqn.(2), and an apparently dispersive phase shift. However, the length of the constriction is now wavelength (velocity) dependent, $l = l(\lambda)$, and no longer a constant as in the static case.